%% file: compass_3pi2004_v11.tex
\documentclass[12pt]{cernart}

\ifx\pdfoutput\undefined
  \usepackage{graphicx}
\else
  \usepackage[pdftex]{graphicx}
  \usepackage{epstopdf}
\fi

\input myunits.sty
\input mymath.sty

\usepackage{lineno}

\begin {document}
\dimen\footins=\textheight

\begin{titlepage}
\docnum{\vbox{%
CERN--PH--EP/2009--018\\
22. September 2009\\
updated 29. April 2010}}
\vspace{1cm}

\title{Observation of a $J^{PC}=1^{-+}$ exotic resonance in
  diffractive dissociation of $190\,\GeV/c$ $\pi^-$ into 
  $\pi^-\pi^-\pi^+$}

\vspace*{0.5cm}
\collaboration{The COMPASS Collaboration}

\vspace{2cm}
\begin{abstract}
The COMPASS experiment at the CERN SPS has
studied the 
diffractive dissociation of negative pions into the 
$\pi^-\pi^-\pi^+$ final state using a
$190\,\GeV/c$ pion beam hitting a lead target.  
A partial wave
analysis has been performed on a sample of $420\,000$ events 
taken at values of the squared 4-momentum transfer $t'$ between
$0.1$ and $1\,\GeV^2/c^2$. The well-known resonances 
$a_1(1260)$,  
$a_2(1320)$, and   
$\pi_2(1670)$ are clearly observed. In addition, the
data show a significant natural parity exchange production of
a resonance with spin-exotic quantum numbers $J^{PC}=1^{-+}$ at 
$1.66\,\GeV/c^2$ decaying to $\rho\pi$.   
The resonant nature of this wave is evident from the mass-dependent phase
differences 
to the $J^{PC}=2^{-+}$ and $1^{++}$
waves. From a mass-dependent fit a resonance mass of
$(1660\pm 10^{+0}_{-64})\,\MeV/c^2$ and a width of 
$(269\pm21^{+42}_{-64})\,\MeV/c^2$ are deduced, with an intensity of
$(1.7\pm0.2)\%$ of the total intensity. 

\vspace*{60pt}
\end{abstract}

\vfill
\submitted{submitted to Physical Review Letters}

\begin{Authlist}
{\large  The COMPASS Collaboration}\\[\baselineskip]
{\small
\input{auth_cern.tex}}
\end{Authlist}
\mbox{}\\
\input{inst_cern.tex}\end{titlepage}



In the SU(3)$_\mathrm{flavor}$ constituent quark model, light mesons are
described as bound
states of a quark $q$ and an antiquark $\overline{q}'$ with quark flavors
$u,d,s$. 
Mesons are classified in $J^{PC}$ multiplets, with 
the total angular momentum $J$, the parity $P$, and the 
particle-antiparticle conjugation parity $C$, 
which is defined through  
the neutral flavorless members of a given multiplet. 
The isospin $I$ and the $G$-parity further characterize mesons
containing light quarks. 
%
%
In the quark model, $P$, $C$ and $G$ are given by 
\begin{equation}
  \label{eq:meson.qn}
  P=(-1)^{L+1},\quad C=(-1)^{L+S},\quad G=(-1)^{I+L+S},
\end{equation}
where $L$ is the relative orbital angular momentum of $q$ and $\overline{q}'$, 
and $S$ the total intrinsic spin of the $q\overline{q}'$ pair, with
$S=0,1$. 
The constituent quark model has been quite successful
in explaining many of the properties of mesons as well as, to a large extent,
the observed
meson spectrum, even though it makes no assumptions concerning the 
nature of the 
binding force, except that hadrons are postulated to be color-singlet
states. In Quantum Chromodynamics (QCD),  
the interaction
between colored  
quarks is described by the exchange of gluons which carry color themselves.
Owing to this particular structure of QCD, color-singlet mesons can
be formed not only by constituent quarks, but also by other
configurations like four-quark objects or 
gluonic excitations. 
These non-$q\overline{q}'$ configurations, however, will mix with ordinary
$q\overline{q}'$ states 
with the same $J^{PC}$, making
it difficult to disentangle the contribution of each configuration. 
The observation of exotic states with quantum numbers not allowed in the
simple quark model, e.g.\
$J^{PC}=0^{--},\,0^{+-},\,1^{-+},\ldots$, would
give clear evidence 
that quark-gluon configurations beyond the quark model, as allowed by
QCD, are realized in nature. 

The lowest-lying hybrid, i.\ e.\ a system consisting of a color octet
$q\overline{q}'$ pair neutralized in color by a
gluonic excitation,  
is expected \cite{Juge:2003qd} to
have exotic quantum 
numbers $J^{PC}=1^{-+}$, and thus will not mix with
ordinary mesons. Its mass is predicted in the region $1.3 -
2.2\,\GeV/c^2$.
The systematics of hybrid meson production and decay has been
worked out in the flux-tube model \cite{Close:1994hc}. 
There are 
three 
experimental candidates for a light $1^{-+}$ hybrid. The
$\pi_1(1400)$ was observed by E852 \cite{Thompson:1997bs} and by
VES \cite{Dorofeev:2001xu} in the reaction   
$\pi^- N\rightarrow \eta\pi^- N$, and 
by 
Crystal Barrel \cite{Abele:1998gn,Abele:1999tf}
in $\overline{p}n\rightarrow\pi^-\pi^0\eta$ and
$\overline{p}p\rightarrow 2\pi^0\eta$ Dalitz plot
analyses.  
Another $1^{-+}$ state, the $\pi_1(1600)$, decaying into
$\rho\pi$ \cite{Adams:1998ff,Chung:2002pu,Khokhlov:2000tk},
$\eta'\pi$ \cite{Beladidze:1993km,Ivanov:2001rv},  
$f_1(1285)\pi$ \cite{Kuhn:2004en,Amelin:2005ry}, and
$b_1(1235)\pi$ \cite{Amelin:2005ry,Lu:2004yn} was observed in
peripheral $\pi^- p$ 
interactions in E852 and VES, and confirmed in
$\overline{p}p\rightarrow b_1\pi\pi$ \cite{Baker:2003jh}. 
The resonant nature of both states, however, is
still heavily disputed in the 
community \cite{Dorofeev:2001xu,Amelin:2005ry}. 
In a different analysis of a larger data set of 
E852 no evidence for an exotic resonance at $1.6\,\GeV/c^2$ in the
$3\pi$ final state was 
found \cite{Dzierba:2005jg}. 
A third exotic state, $\pi_1(2000)$, decaying to $f_1\pi$ and $b_1\pi$, was
seen in only one experiment \cite{Kuhn:2004en,Lu:2004yn}.

In order to shed new light on these questions, 
the COMPASS collaboration, operating a large-acceptance and high-resolution
spectrometer \cite{Abbon:2007pq} situated at the CERN Super Proton
Synchrotron (SPS), is gathering
high-statistics event samples of 
diffractive 
reactions of hadronic probes into final states containing both
charged and neutral particles. 
Diffractive dissociation is a reaction of the type  
$a+b\rightarrow c+d$ with $c\rightarrow
1+2+\cdots+n$, 
where $a$ is the incoming beam 
particle, $b$ the target, $c$ the diffractively produced object
decaying into $n$ particles, and $d$
the target recoil particle, with 4-momenta $p_a\ldots p_d$, respectively. 
The production kinematics is described by 
two variables: $s$  and $t' = \vert t \vert - \vert t \vert_\mathrm{min}$, 
where
$s = (p_{a}+p_{b})^2$  is the square of the total center of mass energy, 
$t = (p_{a} - p_{c})^2$ is the square of the four momentum transferred from the
incoming beam to the outgoing system $c$, and $\vert t \vert_\mathrm{min}$ is 
the minimum value of $\vert t \vert$ which
is allowed by kinematics for a given mass $m_c$.  

\begin{figure}[tbp]
  \centering
  \includegraphics[width=0.45\textwidth]{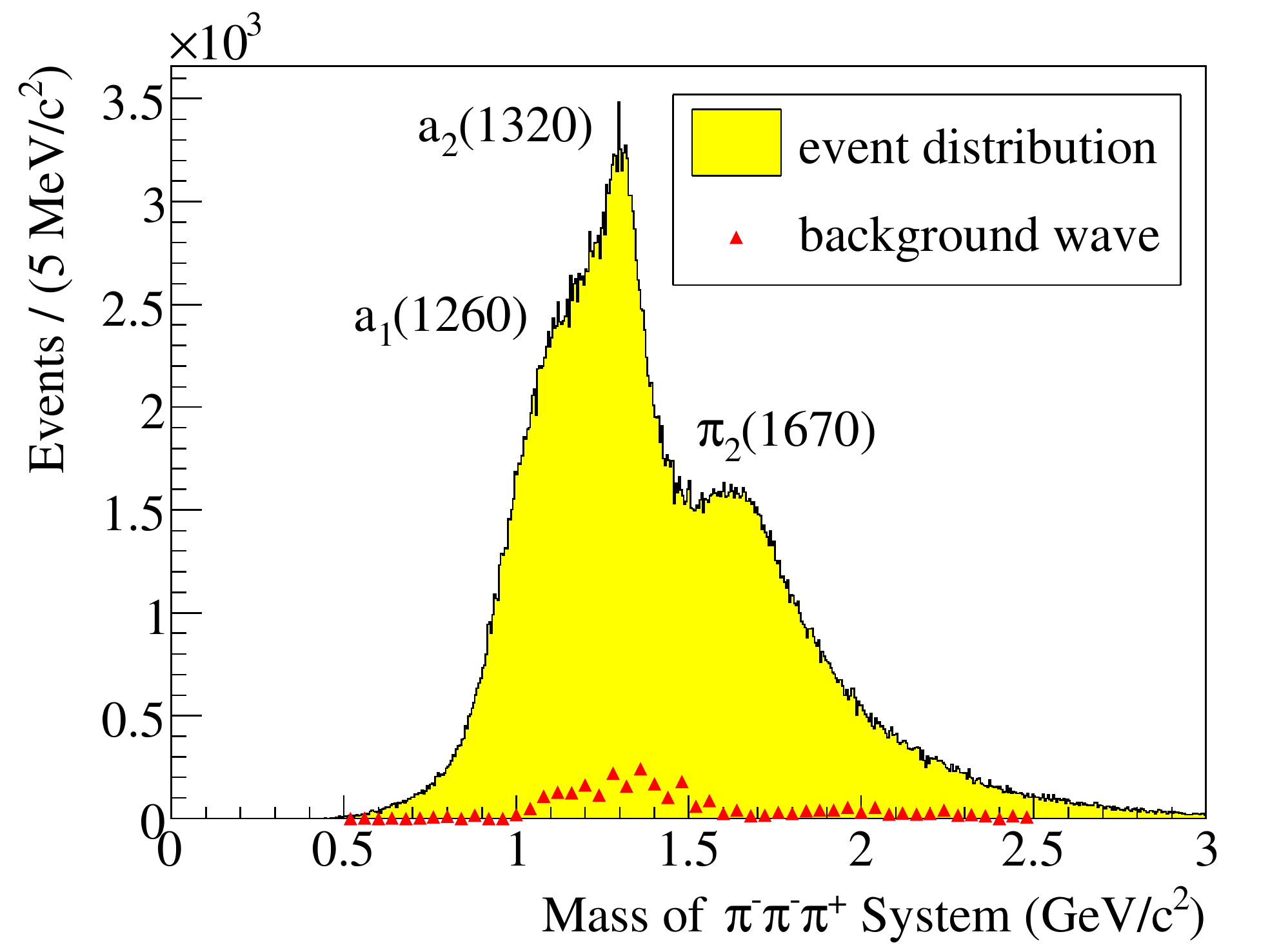}
  \hfill
  \caption{Invariant mass of the $3\pi$ system for
    $0.1\,\GeV^2/c^2<t'<1.0\,\GeV^2/c^2$ (histogram), and 
    intensity of the background
    wave with a flat distribution in
    3-body phase space (triangles), obtained from a partial wave
    analysis in $40\,\MeV/c^2$ bins of the $3\pi$ mass and rescaled
    to the binning of the histogram. Both the
    invariant mass spectrum and the background distribution are not
    acceptance corrected.}  
  \label{fig:3pi.mass}
\end{figure}

First studies of diffractive reactions of 
$190\,\GeV/c$ $\pi^-$ on a $3\,\mm$ lead 
target were carried out by COMPASS in 2004. The $\pi^- \pi^-
\pi^+$ final state was chosen because the disputed $\pi_1(1600)$ meson
with exotic $J^{PC}$ had previously been reported in
this channel.  
The trigger selected events with one incoming particle and at least
two outgoing charged particles. In the
offline analysis, a primary 
vertex inside the target with 3 outgoing charged particles is
required. Since the recoil particle was not 
detected, the following procedure is applied in order to select
exclusive events. 
The beam energy $E_a$ is 
very well approximated by the measured 
total energy $E_c$ of the $3\pi$ system 
with a small correction arising from the target recoil, which can be
calculated from the measured 
scattering angle $\theta=\angle(\vec{p}_a,\vec{p}_c)$,  
assuming that the target particle remained intact throughout
the scattering process. 
Then an exclusivity cut is
applied, requiring $E_a$ to be within $\pm 4\,\GeV$ of the mean 
beam energy. 
Events with 
a wide range of $t'$ from zero up to a few $\GeV^2/c^2$
were recorded. 
For the analysis presented in this letter 
we restrict ourselves to the range 
where candidates for spin exotic 
states have been reported in the past: $0.1\,\GeV^2/c^2< t' <1.0\,\GeV^2/c^2$,
far beyond the region of 
coherent scattering on the Pb nucleus.   
Figure~\ref{fig:3pi.mass} shows the invariant mass
of the corresponding events.
In our sample of $420\,000$
events in the mass range between $0.5$ and $2.5\,\GeV/c^2$, the
well-known resonances $a_1(1260)$, $a_2(1320)$, and 
$\pi_2(1670)$ are 
clearly visible in the $3\pi$ mass spectrum. 

A partial wave analysis (PWA) of this data set was performed using a
program which was originally developed at Illinois \cite{Ascoli:1970xi}, and
later modified at 
Protvino and Munich. 
An independent cross-check of the results was performed using a different PWA program
developed at Brookhaven \cite{Cummings:1997a} and adapted for COMPASS
\cite{Neubert:2009a}. 
At high $\sqrt{s}$, the reaction can be assumed to proceed via $t$-channel
Reggeon exchange, thus justifying the factorization 
of the total cross section into a resonance and a recoil vertex
without final state interaction. 
The exchanged Reggeon may excite the incident pion
($J^{P}=0^{-}$) 
to a state $X$ with different
$J^{P}$, limited only by conservation laws for strong
interactions. For the $(3\pi)^-$ final state $I\geq 1$; we assume
$I=1$ since no flavor-exotic mesons have been found. Since in addition
$G=-1$ for a system with an odd number of pions, $C=+1$ follows from 
eq.~\ref{eq:meson.qn}.  
We take the phenomenological approach of the isobar model, in which
all multi-particle final states 
can be described by sequential two-body decays into intermediate resonances
(isobars), which eventually decay into the final state observed in the
experiment.  
All known isovector and isoscalar $\pi\pi$
resonances have been included in our fit: 
$(\pi\pi)_S$ (comprising the
broad 
$\sigma(600)$ and $f_0(1370)$), $\rho(770)$, $f_0(980)$, $f_2(1270)$,
and $\rho_3(1690)$ \cite{Chung:2002pu}.  
It is possible that there exists a
direct 3-body decay into $(3\pi)^-$ without an intermediate di-pion resonance;
in the isobar model, such a decay mode without angular correlations is 
represented by $\sigma(600)+\pi^-$ with $L=0$ and $J^P=0^-$. 
Possible complications to the isobar model from unitarity constraints are 
not an issue here; such effects enter in the formulation of the 
model only when all possible decay modes are simultaneously fit, which may
include 
the final states containing $\pi^0$, $\eta$, $\eta'$, $\omega$ , $K\bar K$ or
$N\bar N$. 
The spin-parity composition of the excited state $X$ is studied in the 
Gottfried-Jackson frame, which is the center of mass frame of $X$ with
the $z$-axis along the beam direction, and the $y$-axis perpendicular
to the production plane, formed by the momentum vectors of the target
and the recoil particle. 

The PWA is done in two steps. In the first step, a 
fit of the probability density in $3\pi$ phase space is performed in
$40\,\MeV/c^2$  
bins of the $3\pi$ invariant mass $m$ (fit in mass bins). 
No dependence of the production strength for a given wave on the mass
of the
$3\pi$ system is introduced at this point:
\begin{equation}
  \label{eq:sigma_indep}
  \sigma_\mathrm{indep}(\tau,m,t')=\sum_{\epsilon=\pm 1}\sum_{r=1}^{N_r} 
  \left|\sum_{i} {T_{ir}^\epsilon
      f_i^\epsilon(t')\psi_i^\epsilon(\tau,m)}/N_i^\epsilon(m)\right|^2,\quad 
  N_i^\epsilon(m)=  
  {\sqrt{\int\left|\psi_i^\epsilon(\tau',m)\right|^2\diff{\tau'}}}\quad. 
\end{equation}
Here, $T_{ir}^\epsilon$ are the production amplitudes and $\psi_i^\epsilon$
the decay amplitudes, the indices $i$ and
$\epsilon$ denoting 
different partial waves, characterized by a set of
quantum numbers  
$J^{PC}M^\epsilon[\mathrm{isobar}]L$; 
$M$ is the absolute value of the spin projection onto the
$z$-axis; $\epsilon$ is the
reflectivity \cite{Chung:1974fq}, which describes the symmetry under a
reflection through the 
production plane, and which is 
defined such that it 
corresponds to the naturality of
the exchanged Regge trajectory; $L$ is the orbital angular momentum
between the isobar and the bachelor pion. 
The different $t'$ dependence of the cross section for
$M=0$ and $M=1$ states is taken into 
account by including different functions of $t'$, $f_i^\epsilon(t')\propto
\exp{(-bt')}$ ($M=0$) and 
$f_i^\epsilon(t')\propto t'\exp{(-bt')}$ ($M=1$),  
where the slope $b$ has been obtained from the data by first 
making fits in slices of $t'$.  
The 3-body decay amplitudes $\psi_i^\epsilon$ 
are constructed using non-relativistic Zemach tensors
\cite{Chung:1971ri}. They are properly Bose-symmetrized to take into
account the combinatorics due to the spin-0 nature of the final state
pions. They depend on the set of five 
parameters $\tau$ specifying the 3-body decay kinematics, but 
do not contain any free parameters. 
The normalization factors $N_i^\epsilon(m)$ contain angular-momentum  
barrier factors and quasi-2-body phase space factors, taking into
account the non-zero widths of isobars. 
Dividing each decay amplitude by $N_i^\epsilon(m)$ compensates its dependence on
the mass inside each mass 
bin. 
Equation~\ref{eq:sigma_indep} includes a coherent sum over waves with
different $J^{PC}M$, allowing them to interfere.   
It also contains two non-coherent sums over the reflectivity $\epsilon$ and
the rank $N_r$ \cite{Chung:1974fq}. 
Assuming that the recoiling target particle is a  
nucleon, and
neglecting nuclear effects, we set
$N_r=2$, 
corresponding to a mixture of helicity-flip and helicity-non-flip
processes at the baryon vertex. 
A total of 42 partial waves are included in the first step of the fit. It
comprises the non-exotic positive-reflectivity waves with $J^{PC}=0^{-+}$
($M=0$), $1^{++},2^{-+},3^{++},4^{-+}$ ($M=0,1$), $2^{++},4^{++}$
($M=1$), the exotic $1^{-+}$ ($M=1$), and the negative-reflectivity 
waves $1^{-+},2^{++}$ ($M=0,1$), $1^{++},2^{-+}$ ($M=1$), 
taking into account all relevant known decay modes into the isobars listed above.   
The fit also contains a background wave, characterized by a 
uniform distribution in 3-body phase space, which is added
incoherently to the other waves.  
In the fit the elements of the spin-density matrix, given by
$\rho_{ij}^\epsilon=\sum\nolimits_r T_{ir}^\epsilon T_{jr}^{\epsilon\ast}$,  
are determined simultaneously for all 42 waves in a given mass bin 
using an extended maximum likelihood method. 
The diagonal elements of the spin-density matrix are the intensities
of the corresponding waves, while the off-diagonal elements determine
the phase differences between two waves. 
There is no a-priori
assumption on a resonant behavior of a 
given wave at this first step.   
The fit also takes into
account 
the experimental
acceptance of the spectrometer, calculated from a phase-space Monte Carlo
simulation of the 
apparatus. It is worth stressing that COMPASS has an excellently
uniform   
acceptance for diffractively produced 3$\pi$ events of the order of $60\%$
over the whole phase space. 
In order to verify that indeed the global maximum has been found by the fit,
up to $100$
attempts with randomly chosen start parameters are performed for each mass
bin. If multiple solutions are found
within one unit of log likelihood, 
the average of the two extreme 
solutions for each parameter is used. The difference of
the two  
extreme solutions is added linearly to the statistical error of the
best fit for each parameter. 

In the second step of the PWA a $\chi^2$ fit of the 
spin-density matrix elements obtained for each mass bin in the first
step is performed in the mass range from $0.8$ to $2.32\,\GeV/c^2$,
taking 
into account the mass dependence of the produced 
resonances (mass-dependent fit). 
The elements of the spin density matrix are expressed as 
$\rho_{ij}^\epsilon=\sum\nolimits_r
A_{ir}^\epsilon(m)A_{jr}^{\epsilon\ast}(m)$ with amplitudes 
$A_{ir}^\epsilon(m)=\sum\nolimits_k
C_{ikr}^\epsilon\,BW_k(m)\,N_i^\epsilon(m)$, with 
$BW_k(m)$ denoting relativistic Breit-Wigner functions (with
constant or dynamic 
widths depending on whether branching ratios of the corresponding
resonance are known) or, where
required by the fit, a coherent background reflecting non-resonant
production of the corresponding partial wave, e.g.\ via the Deck effect
\cite{Deck:1964hm}. For the latter, an empirical parameterization
consisting of 
a simple exponential, $\exp(-\alpha p^2)$, with $p$ being the break-up
momentum for the 2-body decay of the produced resonance and $\alpha$ a fit 
parameter   
is used.   
In the mass-dependent fit the complex production amplitudes
$C_{ikr}^\epsilon$, and the parameters of $BW_k(m)$ are determined    
for a subset of six waves, the
selected waves showing either significant amplitudes or rapid relative
phase changes in the $1.7\,\GeV/c^2$ mass range:   
$0^{-+}0^+\,f_0(980)\pi\,S$,
$1^{++}0^+\,\rho\pi\,S$,
$2^{-+}0^+\,f_2\pi\,S$,
$2^{++}1^+\,\rho\pi\,D$,
$4^{++}1^+\,\rho\pi\,G$,  
and the exotic 
$1^{-+}1^+\,\rho\pi\,P$. 

\begin{figure*}[tp]
  \centering
  \includegraphics[width=0.45\textwidth]{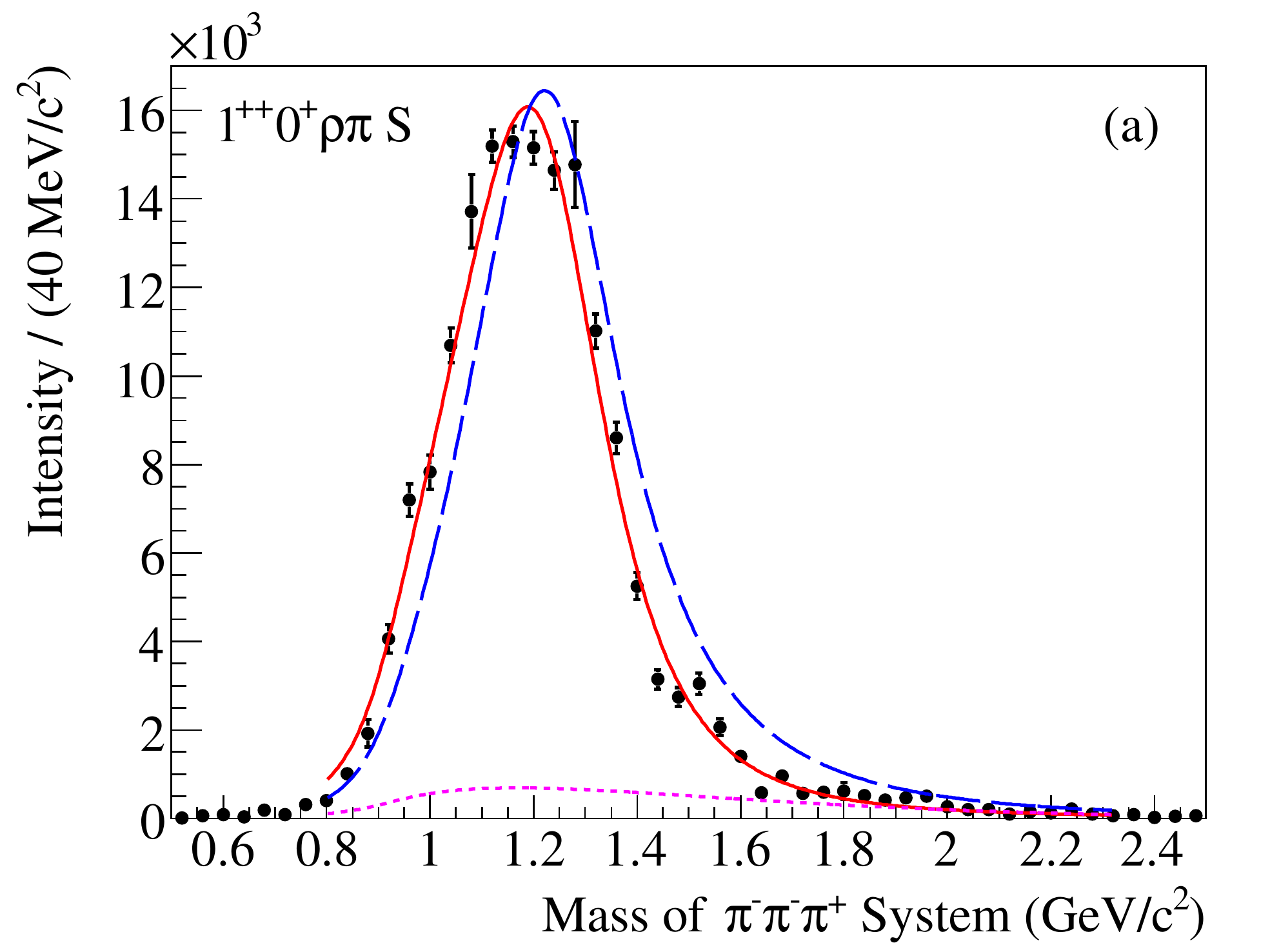}
  \hfill
  \includegraphics[width=0.45\textwidth]{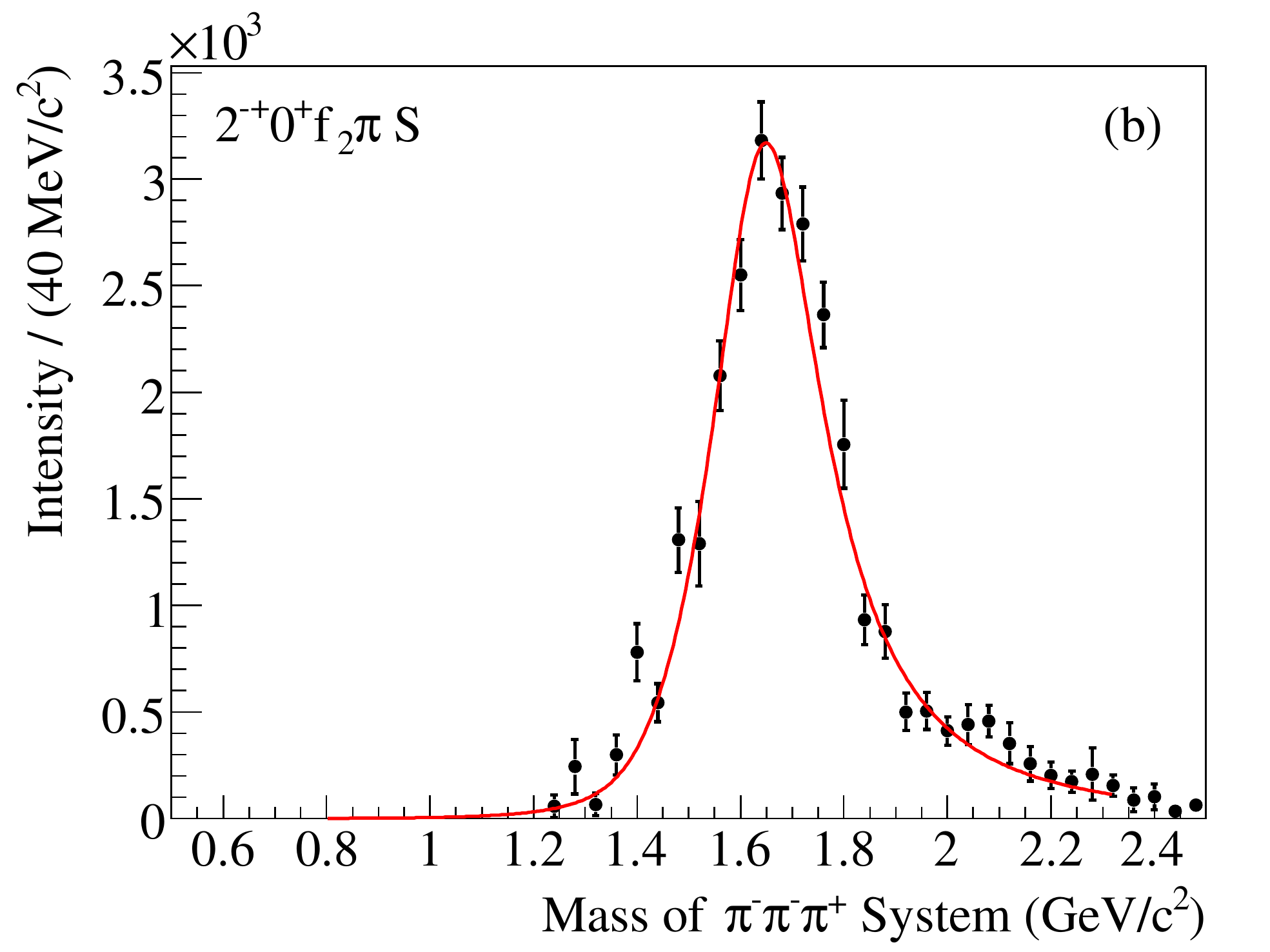}
  \hfill\\
  \includegraphics[width=0.45\textwidth]{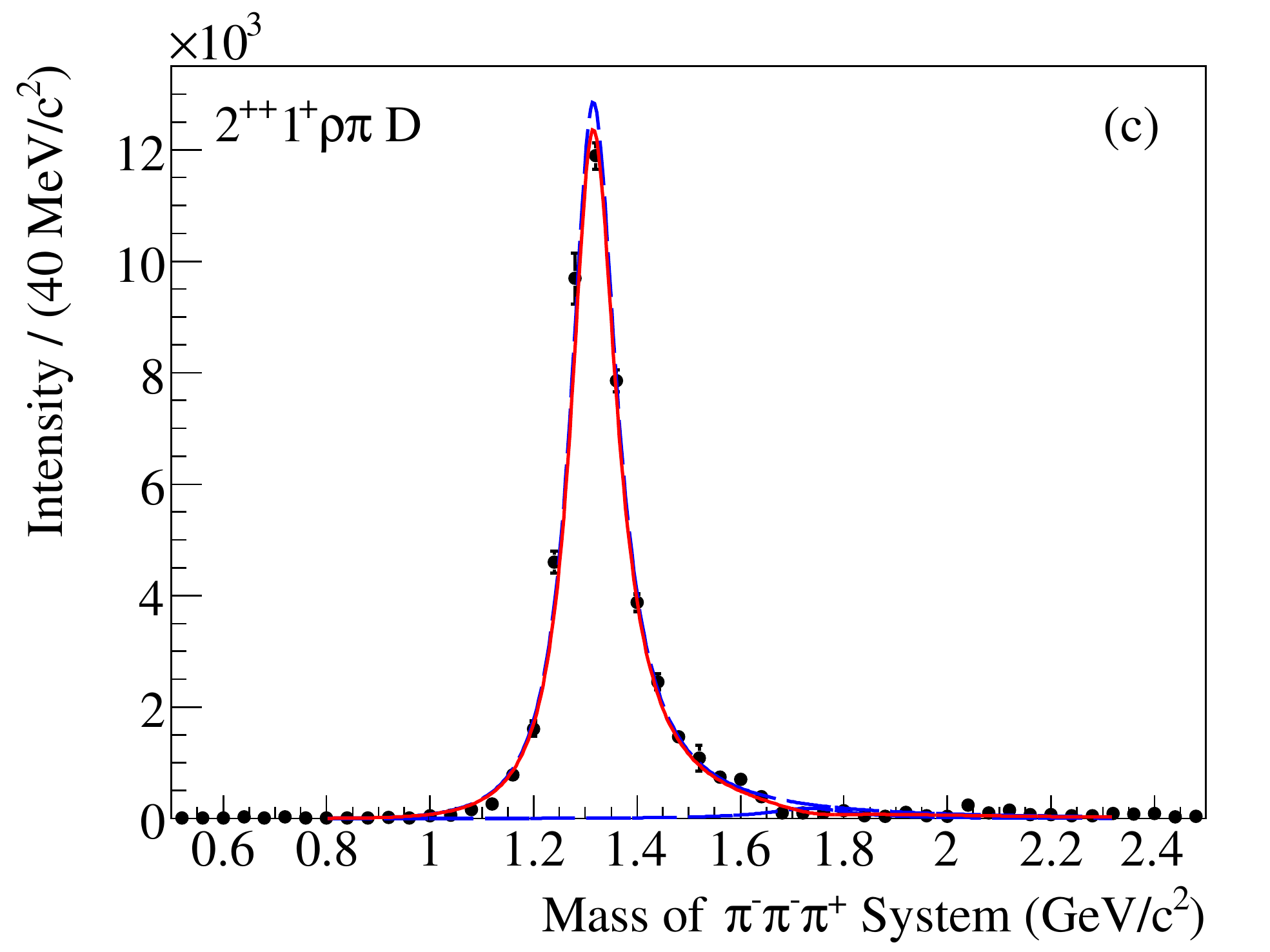}
  \hfill
  \includegraphics[width=0.45\textwidth]{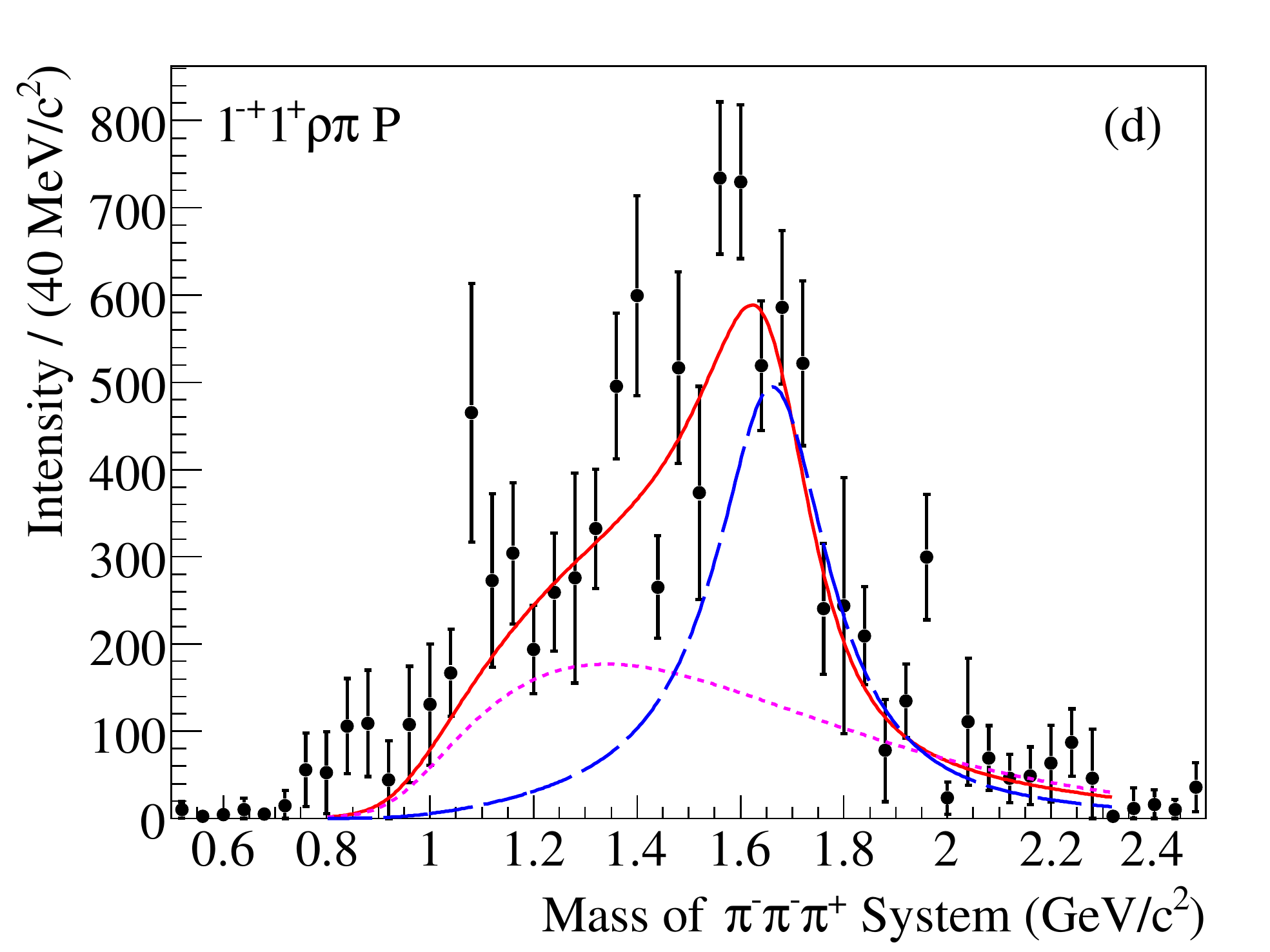}
  \hfill
  \caption{Intensities of major waves
  $1^{++}0^+\,\rho\pi\,S$ (a),  
    $2^{-+}0^+\,f_2\pi\,S$ (b), and $2^{++}1^+\,\rho\pi\,D$ (c), as
    well as the intensity of the exotic wave 
    $1^{-+}1^+\,\rho\pi\,P$ (d), as determined in the fit in mass bins
    (data points with error bars). The lines represent the result of
    the mass-dependent fit (see text).}  
  \label{fig:3pi.intensities}
\end{figure*}

\begin{figure*}[tp]
  \centering
  \includegraphics[width=0.45\textwidth]{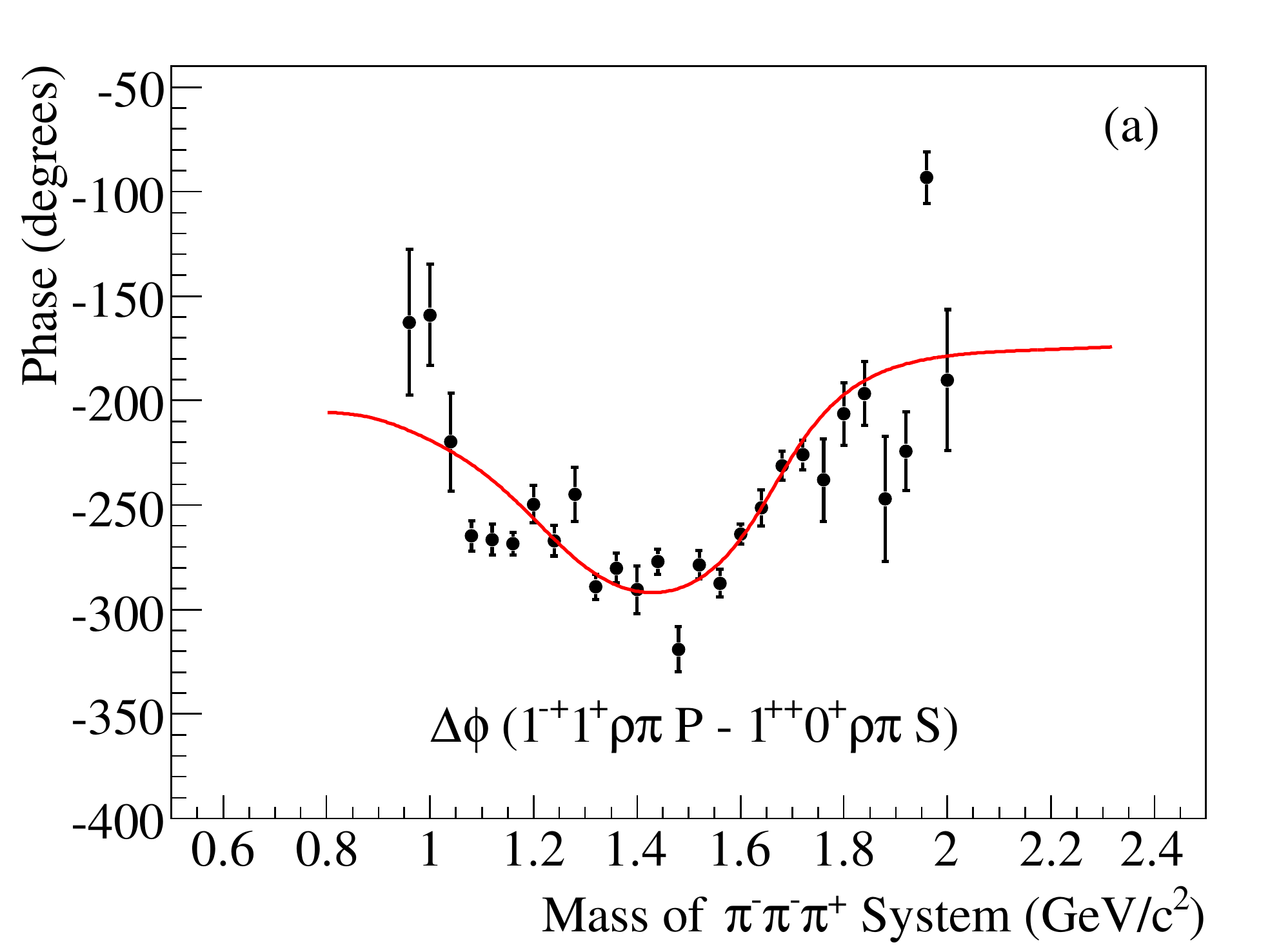}
  \hfill
  \includegraphics[width=0.45\textwidth]{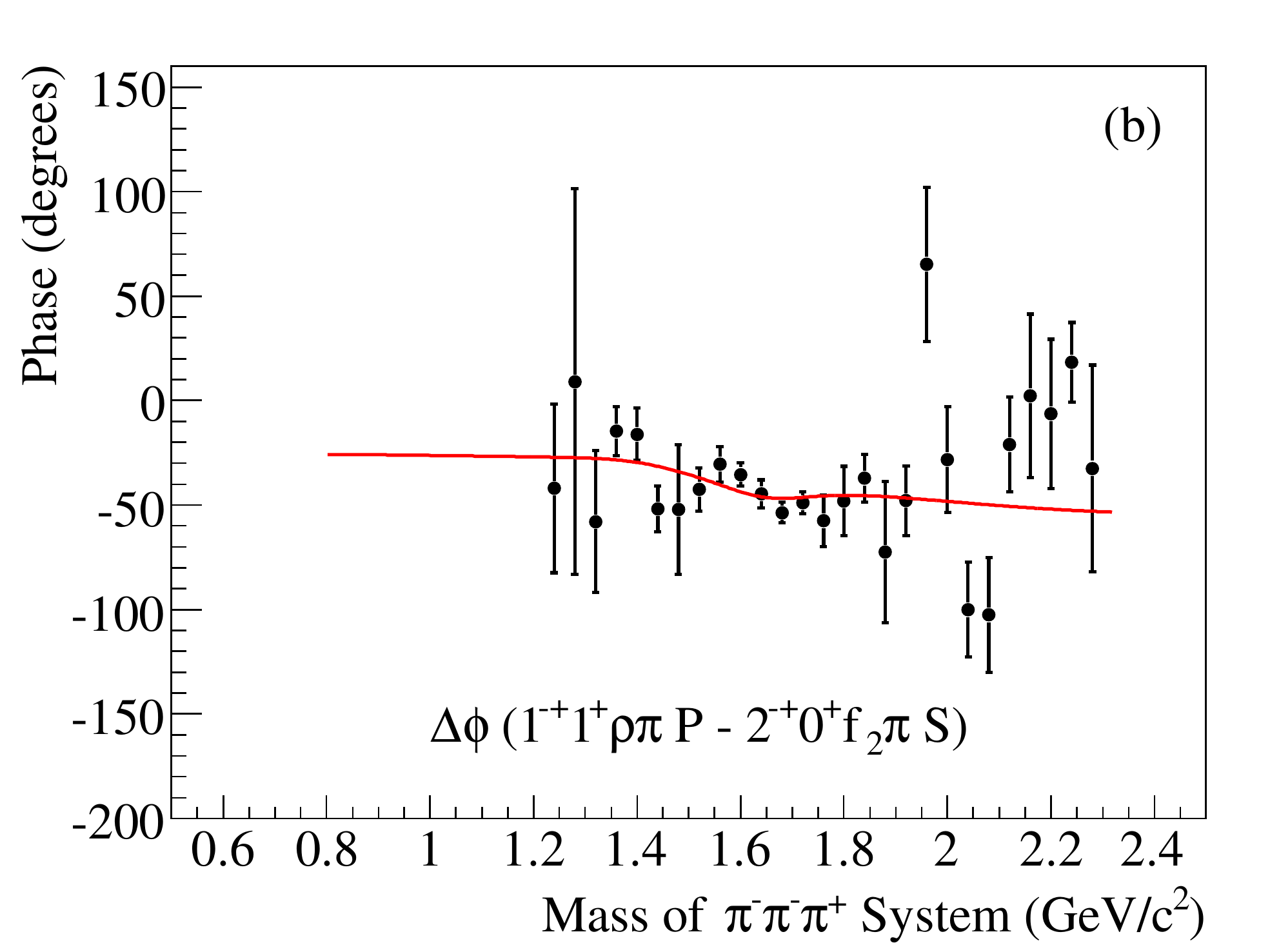}
  \hfill
  \caption{Phase differences of the exotic $1^{-+}1^+\,\rho\pi\,P$ wave to the
    $1^{++}0^+\,\rho\pi\,S$ (a) 
    and the $2^{-+}0^+\,f_2\pi\,S$ (b) waves. The data points
    represent the result of the fit in mass bins, the lines are the
    result of the mass-dependent fit.} 
  \label{fig:3pi.phases}
\end{figure*}

The intensity of the background
wave resulting from the fit in mass bins is included in
Figure~\ref{fig:3pi.mass} 
(triangles).  
Figures~\ref{fig:3pi.intensities}(a)-(c) show the acceptance-corrected 
intensities of the three most prominent waves $1^{++}0^+\,\rho\pi\,S$,
$2^{-+}0^+\,f_2\pi\,S$, and $2^{++}1^+\,\rho\pi\,D$, respectively, as
determined from the fit in mass bins (data points with error bars). 
The 
data also show a significant natural parity exchange production of 
a wave 
with spin-exotic quantum numbers $J^{PC}=1^{-+}$ at 
$1.66\,\GeV/c^2$ decaying to $\rho\pi$ ($P$ wave), presented in
Fig.~\ref{fig:3pi.intensities}(d).  
The resonant nature of the exotic wave is evident from its
phase 
differences to the dominant $1^{++}0^+\,\rho\pi\,S$
and $2^{-+}0^+\,f_2\pi\,S$ waves, shown in
Figs.~\ref{fig:3pi.phases}(a) and (b), respectively (data
points). 
For the latter, shown in
Fig.~\ref{fig:3pi.phases}(b), no significant change in 
the phase difference between $1.4$ and $1.9\,\GeV/c^2$ is observed, which is
attributed 
to the fact that there are two resonances, $\pi_1(1600)$ and $\pi_2(1670)$,
with very similar masses and widths, causing the relative phase difference 
to be almost constant. In contrast to this the phase difference to the $1^{++}$
wave, shown in 
Fig.~\ref{fig:3pi.phases}(a), 
clearly shows an increase around $1.7\,\GeV/c^2$. As the
$a_1(1260)$ is no longer resonating at this mass, this observation can
be regarded as an independent verification of the resonating nature of
the $1^{-+}$ wave.  

The solid lines in 
Fig.~\ref{fig:3pi.intensities} show the total intensity from the
mass-dependent fit 
for the corresponding waves. 
For the 
$1^{++}0^+\,\rho\pi\,S$ wave shown in
Fig.~\ref{fig:3pi.intensities}(a) it is well known that there is a
significant contribution of 
non-resonant production through the Deck effect \cite{Daum:1980ay},
indicated by the dotted line. Its interference with the $a_1(1260)$  
(dashed line) shifts the peak in the data to a slightly lower value
than the peak position of the resonance. 
The $2^{-+}0^+\,f_2\pi\,S$ wave shown in
Fig.~\ref{fig:3pi.intensities}(b) is well described by a single
resonance, the 
$\pi_2(1670)$. 
The $2^{++}1^+\,\rho\pi\,D$ wave displayed in
Fig.~\ref{fig:3pi.intensities}(c) is dominated by the $a_2(1320)$
with a small contribution from the $a_2(1700)$, whose parameters have
been fixed to PDG values \cite{Yao:2006px} because of the
limited statistics.
The intensity of
the exotic $1^{-+}1^+\,\rho\pi\,P$, shown in
Fig.~\ref{fig:3pi.intensities}(d),  
is well described by 
a Breit-Wigner resonance with constant width at $1.66\,\GeV/c^2$
(dashed line), which we 
interpret as the $\pi_1(1600)$, and a non-resonant 
background (dotted line) at lower masses. The resonant component of
the exotic wave is strongly constrained by the mass-dependent phase 
differences to the $1^{++}0^+\,\rho\pi\,S$ and the 
$2^{-+}0^+\,f_2\pi\,S$ waves, which are well reproduced in the
mass-dependent fit (solid lines in Fig.~\ref{fig:3pi.phases}). 
The parameters deduced for the masses, widths and intensities of the resonances
included in the mass-dependent fit are given in
Table~\ref{tab:3pi.results}, 
where the first
uncertainty corresponds to the statistical error, the second to the
systematic error. The intensities are given for the resonant part of the 
corresponding wave integrated over the mass range from $0.8$ to
$2.32\,\GeV/c^2$, and are normalized to the total intensity from the 
mass-dependent fit, corresponding to $38.7(2)\%$ of the
acceptance-corrected data sample in the same mass range.  
The intensity of the $\pi_1(1600)$ is found to be $(1.7\pm0.2)\%$
of the total intensity, while the non-resonant contribution is
$(1.5\pm0.2)\%$, with a linear correlation coefficient between the two
intensities of $15\%$.  
\begin{table*}[tbp]
  \scriptsize
  
  \centering
  \begin{tabular}{lllllll}
    \hline\hline
    Resonance & Mass & Width &
    Intensity & Channel & Mass \cite{Amsler:2008zzb} & Width \cite{Amsler:2008zzb}\\ 
              & ($\MeV/c^2$) & ($\MeV/c^2$) &
    ($\%$)    & $J^{PC}M^\epsilon[\mathrm{isobar}]L$ 
              & ($\MeV/c^2$) & ($\MeV/c^2$) \\ \hline 
    $a_1(1260)$ & $1255\pm 6^{+7}_{-17}$ & $367\pm 9^{+28}_{-25}$ &
    $67\pm 3^{+4}_{-20}$ & $1^{++}0^+\,\rho\pi\,S$ &
    $1230\pm40$ & $250-600$ \\
    $a_2(1320)$ & $1321\pm 1^{+0}_{-7}$  & $110\pm 2^{+2}_{-15}$ &
    $19.2\pm 0.6^{+0.3}_{-2.2}$ & $2^{++}1^+\,\rho\pi\,D$ & 
    $1318.3\pm 0.6$ & $107\pm 5$ \\
    $\pi_1(1600)$ & $1660\pm 10^{+0}_{-64}$ & $269\pm 21^{+42}_{-64}$ &
    $1.7\pm 0.2^{+0.9}_{-0.1}$ & $1^{-+}1^+\,\rho\pi\,P$ & 
    $1662^{+15}_{-11}$ & $234\pm 50$ \\
    $\pi_2(1670)$ & $1658\pm 3^{+24}_{-8}$ & $271\pm 9^{+22}_{-24}$ &
    $10.0\pm 0.4^{+0.7}_{-0.7}$ & $2^{-+}0^+\,f_2\pi\,S$ & 
    $1672.4\pm 3.2$ & $259\pm 9$ \\
    $\pi(1800)$ & $1785\pm 9^{+12}_{-6}$ & $208\pm 22^{+21}_{-37}$ &
    $0.8\pm 0.1^{+0.3}_{-0.1}$ & $0^{-+}0^+\,f_0\pi\,S$ & 
    $1816\pm 14$ & $208\pm 12$ \\
    $a_4(2040)$ & $1885\pm 13^{+50}_{-2}$ & $294\pm 25^{+46}_{-19}$ &
    $1.0\pm 0.3^{+0.1}_{-0.1}$ & $4^{++}1^+\,\rho\pi\,G$ & 
    $2001\pm 10$ & $313\pm 31$ \\ 
    \hline\hline 
  \end{tabular}
  \caption{Resonance masses, total widths, and intensities for the specified decay
    channel of the six waves included in
    the mass-dependent fit to the data. The first uncertainty
    corresponds to the statistical error, the asymmetric second one to the systematic
    error. The last two columns give the corresponding PDG values
      \cite{Amsler:2008zzb}.}   
  \label{tab:3pi.results}
  \normalsize
\end{table*}
The dominance of natural- over unnatural-parity exchange is more
pronounced than in the BNL case 
at $18\,\GeV/c$ \cite{Chung:2002pu}. This is
possibly due to the 
decreasing contribution of unnatural-parity exchange with an
increasing beam energy, if the natural-parity exchange is mediated
mostly by the Pomeron. 

The systematic errors were estimated from the data by testing the
stability of the 
result with respect to various assumptions made in the analysis,
e.g.\ adding or removing certain waves, varying cuts or initial 
parameters for the fit. One such study concerns the choice of the rank $N_r$ 
used in the PWA. 
Although $N_r=2$ is physically motivated from the
fact that, at high $t'$, incoherent diffraction from individual
nucleons dominates the reaction, fits with $N_r=1$ and $3$ were tried as well. 
The intensity in the background wave relative to the
total acceptance-corrected data sample in the mass range from $0.5$ to
$2.5\,\GeV/c^2$ increases from $5.8\%$ for  
$N_r=2$ to $19\%$ for $N_r=1$, while it drops to $1.2\%$ for
$N_r=3$. At the same time, however, $N_r=3$ was found to cause
larger bin-to-bin fluctuations 
without significantly altering the result.
Given the level of the present
statistics, we therefore conclude that the optimum rank is $N_r=2$. 
In an attempt to account for the low-mass 
shoulder in the intensity of the $1^{-+}1^+\,\rho\pi\,P$ wave we also tried to
include a $\pi_1(1400)$ into the mass-dependent fit, with parameters
fixed to PDG values  
\cite{Yao:2006px}. This reduced the background intensity to a
negligible value and shifted the resonance mass of the $\pi_1(1600)$  
to a slightly smaller value, which is reflected in its systematic
error, but did not 
affect the intensity or the phase differences of 
any of the other waves in the mass-dependent fit. 
Releasing the parameters of the $\pi_1(1400)$, 
however, causes the 
fit to become unstable. This can be attributed to the fact that the
$\pi_1(1400)$, 
if present at 
all, couples only weakly to the $\pi^- \pi^- \pi^+$ final state. 
Another study of the 
sensitivity of the $\pi_1(1600)$ intensity to the functional form
of the background was performed using a background parameterization
without angular-momentum barrier factors, which did not alter the result. 
Other systematic studies included a shift of the $40\,\MeV/c^2$ mass bins by
$20\,\MeV/c^2$, the use of rotation functions with relativistic 
factors \cite{Chung:2007nn} instead of Zemach tensors for the fit in
mass bins, and the 
inclusion of four waves with $M=2$. 
The use of different parameterizations for the $\sigma$ and 
$\rho$ mesons also did not influence the result. 
Performing the fit in mass bins on non-exclusive events, i.e.\
events outside the aforementioned exclusivity cut, no signal is
observed in the $1^{-+}$ wave. 

An incomplete acceptance of the spectrometer, not properly taken into
account in the Monte Carlo simulation, or an incomplete set of waves
may introduce leakage  
of non-exotic waves into the $1^{-+}$ wave. In order to study this effect,
Monte Carlo events were generated using the 
parameters of 16 dominant waves, excluding the $1^{-+}$, which were 
determined in a mass-dependent 
fit, and simulating the decay patterns of the corresponding decay
channels. Performing the same PWA for the Monte Carlo data as for the real
data it was found that the fraction of `fake' intensity in the
observed $1^{-+}$ wave in the Monte Carlo case is less than $5\%$, and thus
negligible. 

In order to test the
significance of the exotic wave, a second fit in mass bins was
performed excluding the exotic wave from the wave set. 
A likelihood ratio test yields a log-likelihood
difference of $47.3$ between the two
fits, averaged over a
mass range of twice the experimental width around the resonance mass
of the $\pi_1(1600)$. For a
difference in the numbers of degrees of freedom of $4$ this confirms
the presence of the exotic wave in the wave set 
with a probability very close to unity.

In conclusion, 
a partial wave analysis of COMPASS data from the diffractive dissociation of
$190\,\GeV/c$ $\pi^-$ on a Pb target into the 
$\pi^-\pi^-\pi^+$ final state at $0.1\,\GeV^2/c^2< t'
<1.0\,\GeV^2/c^2$ was performed. In addition to well-known 
$q\overline{q}'$ 
states, a significant natural parity exchange production of a 
spin-exotic wave with $J^{PC}=1^{-+}$ decaying to $\rho\pi$
is found, with an 
intensity of the resonant part 
corresponding to $(1.7\pm0.2)\%$ of 
the total intensity in the mass-dependent fit. Its mass-dependent phase 
differences to the $J^{PC}=2^{-+}$ and $1^{++}$
waves are consistent
with the highly debated $\pi_1(1600)$ meson. 

We gratefully acknowledge the support of the CERN management and staff
as well as
the skills and efforts of the technicians of the collaborating
institutions. 

\newcommand{\SortNoop}[1]{}

\end{document}

%% file: auth_cern.tex
%
%
M.G.~Alekseev\Iref{turin_i},
V.Yu.~Alexakhin\Iref{dubna},
Yu.~Alexandrov\Iref{moscowlpi},
G.D.~Alexeev\Iref{dubna},
A.~Amoroso\Iref{turin_u},
A.~Austregesilo\IIref{cern}{munichtu},
B.~Bade{\l}ek\Iref{warsaw},
F.~Balestra\Iref{turin_u},
J.~Ball\Iref{saclay},
J.~Barth\Iref{bonnpi}
% ...

%% file: inst_cern.tex
%
%
\Instfoot{bielefeld}{Universit\"at Bielefeld, Fakult\"at f\"ur Physik, 33501 Bielefeld, Germany\Aref{f}}
\Instfoot{bochum}{Universit\"at Bochum, Institut f\"ur Experimentalphysik, 44780 Bochum, Germany\Aref{f}}
\Instfoot{bonniskp}{Universit\"at Bonn, Helmholtz-Institut f\"ur  Strahlen- und Kernphysik, 53115 Bonn, Germany\Aref{f}}
\Instfoot{bonnpi}{Universit\"at Bonn, Physikalisches Institut, 53115 Bonn, Germany\Aref{f}}
\Instfoot{brno}{Institute of Scientific Instruments, AS CR, 61264 Brno, Czech Republic\Aref{g}}
\Instfoot{calcutta}{Matrivani Institute of Experimental Research \& Education, Calcutta-700 030, India\Aref{h}}
\Instfoot{dubna}{Joint Institute for Nuclear Research, 141980 Dubna, Moscow region, Russia}
\Instfoot{erlangen}{Universit\"at Erlangen--N\"urnberg, Physikalisches Institut, 91054 Erlangen, Germany\Aref{f}}
\Instfoot{freiburg}{Universit\"at Freiburg, Physikalisches Institut, 79104 Freiburg, Germany\Aref{f}}
\Instfoot{cern}{CERN, 1211 Geneva 23, Switzerland}
\Instfoot{liberec}{Technical University in Liberec, 46117 Liberec, Czech Republic\Aref{g}}
\Instfoot{lisbon}{LIP, 1000-149 Lisbon, Portugal\Aref{i}}
\Instfoot{mainz}{Universit\"at Mainz, Institut f\"ur Kernphysik, 55099 Mainz, Germany\Aref{f}}
\Instfoot{miyazaki}{University of Miyazaki, Miyazaki 889-2192, Japan\Aref{j}}
\Instfoot{moscowlpi}{Lebedev Physical Institute, 119991 Moscow, Russia}
\Instfoot{munichlmu}{Ludwig-Maximilians-Universit\"at M\"unchen, Department f\"ur Physik, 80799 Munich, Germany\AAref{f}{k}}
\Instfoot{munichtu}{Technische Universit\"at M\"unchen, Physik Department, 85748 Garching, Germany\AAref{f}{k}}
\Instfoot{nagoya}{Nagoya University, 464 Nagoya, Japan\Aref{j}}
\Instfoot{praguecu}{Charles University in Prague, Faculty of Mathematics and Physics, 18000 Prague, Czech Republic\Aref{g}}
\Instfoot{praguectu}{Czech Technical University in Prague, 16636 Prague, Czech Republic\Aref{g}}
\Instfoot{protvino}{State Research Center of the Russian Federation, Institute for High Energy Physics, 142281 Protvino, Russia\Aref{l}}
\Instfoot{saclay}{CEA IRFU/SPhN Saclay, 91191 Gif-sur-Yvette, France}
\Instfoot{telaviv}{Tel Aviv University, School of Physics and Astronomy, 69978 Tel Aviv, Israel\Aref{m}}
\Instfoot{triest_i}{Trieste Section of INFN, 34127 Trieste, Italy}
\Instfoot{triest}{University of Trieste, Department of Physics and Trieste Section of INFN, 34127 Trieste, Italy}
\Instfoot{triestictp}{Abdus Salam ICTP and Trieste Section of INFN, 34127 Trieste, Italy}
\Instfoot{turin_u}{University of Turin, Department of Physics and Torino Section of INFN, 10125 Turin, Italy}
\Instfoot{turin_i}{Torino Section of INFN, 10125 Turin, Italy}
\Instfoot{turin_p}{University of Eastern Piedmont, 1500 Alessandria,  and Torino Section of INFN, 10125 Turin, Italy}
\Instfoot{warsaw}{So{\l}tan Institute for Nuclear Studies and University of Warsaw, 00-681 Warsaw, Poland\Aref{n} }
\Instfoot{warsawtu}{Warsaw University of Technology, Institute of Radioelectronics, 00-665 Warsaw, Poland\Aref{o} }
\Instfoot{yamagata}{Yamagata University, Yamagata, 992-8510 Japan\Aref{j} }
%
%
\Anotfoot{+}{Deceased}
\Anotfoot{a}{Also at IST, Universidade T\'ecnica de Lisboa, Lisbon, Portugal}
\Anotfoot{b}{Also at Chubu University, Kasugai, Aichi, 487-8501 Japan$^{\rm j)}$}
\Anotfoot{c}{Also at KEK, 1-1 Oho, Tsukuba, Ibaraki, 305-0801 Japan}
\Anotfoot{d}{On leave of absence from JINR Dubna}
\Anotfoot{e}{Also at GSI mbH, Planckstr.\ 1, D-64291 Darmstadt, Germany}
\Anotfoot{f}{Supported by the German Bundesministerium f\"ur Bildung und Forschung}
\Anotfoot{g}{Suppported by Czech Republic MEYS grants ME492 and LA242}
\Anotfoot{h}{Supported by SAIL (CSR), Govt.\ of India}
\Anotfoot{i}{Supported by the Portuguese FCT - Funda\c{c}\~{a}o para a
             Ci\^{e}ncia e Tecnologia grants POCTI/FNU/49501/2002 and POCTI/FNU/50192/2003}
\Anotfoot{j}{Supported by the MEXT and the JSPS under the Grants No.18002006, No.20540299 and No.18540281; Daiko Foundation and Yamada Foundation}
\Anotfoot{k}{Supported by the DFG cluster of excellence `Origin and Structure of the Universe' (www.universe-cluster.de)}
\Anotfoot{l}{Supported by CERN-RFBR grant 08-02-91009}
\Anotfoot{m}{Supported by the Israel Science Foundation, founded by the Israel Academy of Sciences and Humanities}
\Anotfoot{n}{Supported by Ministry of Science and Higher Education grant 41/N-CERN/2007/0}
\Anotfoot{o}{Supported by KBN grant nr 134/E-365/SPUB-M/CERN/P-03/DZ299/2000}